\documentclass[aps,prl,twocolumn,superscriptaddress,preprintnumbers,floatfix,longbibliography]{revtex4-1}

\usepackage{dcolumn}
\usepackage{bm}

\usepackage{graphicx}
\usepackage{amsmath}
\usepackage{multirow}
\usepackage{comment}
\usepackage[caption=false]{subfig}
\usepackage{siunitx}
\usepackage{placeins}
\usepackage{color}
\usepackage{standalone}
\usepackage{dcolumn}
\usepackage{bm}
\usepackage{microtype}
\usepackage{etoolbox}
\usepackage{amssymb}
\usepackage{mathrsfs}
\usepackage{accents}
\usepackage[normalem]{ulem}
\usepackage[dvipsnames]{xcolor}
\usepackage[colorlinks,urlcolor=NavyBlue,citecolor=NavyBlue,linkcolor=NavyBlue,pdfusetitle]{hyperref}
\usepackage{gensymb}
\AtBeginDocument{\usepackage{booktabs}}
\graphicspath{{./figs/}}
\usepackage[colorlinks,urlcolor=NavyBlue,citecolor=NavyBlue,linkcolor=NavyBlue,pdfusetitle]{hyperref}

\DeclareMathAlphabet{\mathpzc}{OT1}{pzc}{m}{it}

\newcommand{\sma}[1]{\textcolor{WildStrawberry}{Sizheng: #1}}
\newcommand{\etal}{\textit{et al.\ }}

\usepackage{diagbox}

\begin{document}

\title{Black hole spectroscopy by mode cleaning
}

\newcommand{\Cornell}{\affiliation{Cornell Center for Astrophysics
    and Planetary Science, Cornell University, Ithaca, New York 14853, USA}}
\newcommand\CornellPhys{\affiliation{Department of Physics, Cornell
    University, Ithaca, New York 14853, USA}}
\newcommand\Caltech{\affiliation{TAPIR 350-17, California Institute of
    Technology, 1200 E California Boulevard, Pasadena, CA 91125, USA}}
\newcommand{\AEI}{\affiliation{Max Planck Institute for Gravitational Physics
    (Albert Einstein Institute), Am M\"uhlenberg 1, Potsdam 14476, Germany}} %
\newcommand{\UMassD}{\affiliation{Department of Mathematics,
    Center for Scientific Computing and Visualization Research,
    University of Massachusetts, Dartmouth, MA 02747, USA}}
\newcommand\Olemiss{\affiliation{Department of Physics and Astronomy,
    The University of Mississippi, University, MS 38677, USA}}
\newcommand{\Bham}{\affiliation{School of Physics and Astronomy and Institute
    for Gravitational Wave Astronomy, University of Birmingham, Birmingham, B15
    2TT, UK}}
\newcommand{\ANU}{\affiliation{OzGrav-ANU, Centre for Gravitational Astrophysics, College of Science,
The Australian National University, ACT 2601, Australia}}

\author{Sizheng Ma}
\email{sma@caltech.edu}
\Caltech

\author{Ling Sun}
\ANU

\author{Yanbei Chen}
\Caltech

\hypersetup{pdfauthor={Ma et al.}}

\date{\today}

\begin{abstract}
We formulate a Bayesian framework to analyze ringdown gravitational waves from colliding binary black holes and test the no-hair theorem. The idea hinges on mode cleaning---revealing subdominant oscillation modes by removing
dominant ones using newly proposed {\it rational filters}. By incorporating the filter into Bayesian inference, we construct a likelihood function that depends only on the mass and spin of the remnant black hole (no dependence on mode amplitudes and phases) and implement an efficient pipeline to constrain the remnant mass and spin without Markov chain Monte Carlo (MCMC). 
We test ringdown models by cleaning combinations of different modes and evaluating the consistency between the residual data and pure noise.
The model evidence and Bayes factor are used to demonstrate the presence of a particular mode and to infer the mode starting time.
In addition, we design a hybrid approach to estimate the remnant black hole properties exclusively from a single mode using MCMC after mode cleaning. We apply the framework to GW150914 and demonstrate more definitive evidence of the first overtone by cleaning the fundamental mode. This new framework provides a powerful tool for black hole spectroscopy in future gravitational-wave events. 
\end{abstract}

\maketitle




{\it Introduction.---}
\label{sec:introduction}
The final stage of a binary black hole (BBH) coalescence corresponds to the formation of the remnant BH and its ringing down to a stationary state. The BH linear perturbation theory predicts that the gravitational wave (GW) emitted during this ringdown stage is a superposition of exponentially damped sinusoids called
quasi-normal modes (QNMs) at complex frequencies $\omega_{lmn}$ \cite{Kokkotas:1999bd,Nollert_1999,Cardoso:2016ryw,Berti:2009kk}, labeled by two angular numbers $(l,m)$ and an overtone index $n$. These complex frequencies $\omega_{lmn}$ are fully determined by the mass and spin of the remnant BH due to the no-hair theorem \cite{penrose2002golden,Chrusciel:2012jk,PhysRevLett.26.331,PhysRev.164.1776}, thereby leading to the program of \textit{BH spectroscopy} \cite{PhysRevD.40.3194,Dreyer:2003bv,Berti:2005ys,Berti:2007zu}: the mass and spin of the remnant BH can be measured from the frequency and decay rate of any single QNM in the ringdown regime \cite{PhysRevD.40.3194}; detecting multiple modes allows tests for the no-hair theorem \cite{Dreyer:2003bv,Berti:2005ys,Berti:2007zu}.

A few decades after the conception of BH spectroscopy~\cite{PhysRevD.40.3194,Dreyer:2003bv,Berti:2005ys,Berti:2007zu,Gossan:2011ha,Caudill:2011kv,Meidam:2014jpa,Bhagwat:2016ntk,Berti:2016lat,Baibhav:2017jhs,Maselli:2017kvl,Yang:2017zxs,DaSilvaCosta:2017njq,Baibhav:2018rfk,Carullo:2018sfu,Brito:2018rfr,Nakano:2018vay,Cabero:2019zyt,Bhagwat:2019bwv,Ota:2019bzl,Bustillo:2020buq,JimenezForteza:2020cve,TheLIGOScientific:2016src,Isi:2019aib,Bustillo:2020buq,Isi:2021iql,Cotesta:2022pci,Isi:2022mhy,Finch:2022ynt,Capano:2021etf,Capano:2022zqm}, the first GW detections of BBH mergers \cite{Abbott:2016blz} made by Advanced LIGO \cite{TheLIGOScientific:2014jea,Abbott:2016xvh,TheLIGOScientific:2016agk,Harry:2010zz} allowed the measurement of QNMs from real GW events, such as GW150914 \cite{TheLIGOScientific:2016src,Isi:2019aib,Bustillo:2020buq,Isi:2021iql,Cotesta:2022pci,Isi:2022mhy,Finch:2022ynt} and GW190521 \cite{Capano:2021etf,Capano:2022zqm,Abbott:2020mjq}. 
At late enough times, the fundamental QNM was detected in GW150914 \cite{TheLIGOScientific:2016src,Carullo:2019flw}, with the inferred mass and spin of the remnant BH consistent with the ones obtained from the full inspiral-merger-ringdown (IMR) signal. 
Giesler \etal \cite{Giesler:2019uxc} showed that in a waveform produced by numerical relativity, the ringdown regime starts as early as the time when the strain reaches its peak, if enough overtones are included. 
Motivated by Giesler \etal \cite{Giesler:2019uxc}, Isi \etal \cite{Isi:2019aib} extended the ringdown analysis to earlier times, showing the evidence for the first overtone in GW150914 with a significance of $3.6\sigma$. Studies by Bustillo \etal \cite{Bustillo:2020buq} and Finch \etal \cite{Finch:2022ynt} with different methods also support the existence of the first overtone but with weaker evidence. The conclusion was challenged by Cotesta \etal \cite{Cotesta:2022pci} and then further clarified and discussed by Isi \etal \cite{Isi:2022mhy}. Recently, the existence of the $(l=m=3,n=0)$ mode in the ringdown of GW190521 \cite{Capano:2021etf,Capano:2022zqm,Abbott:2020mjq} is under debate.

A major complication in BH spectroscopy is the non-orthogonality of QNMs, which at low SNRs lead to much ambiguity regarding the existence of modes. Proposals \cite{Finch:2022ynt,Isi:2021iql} have been put forward to solve this issue, including using posteriors of the remnant mass and spin, posteriors of the overtone amplitude, the Bayes factor as a function of analysis start time, and deviations from the overtone frequencies predicted by general relativity. We propose to address this by {\it mode cleaning}, which highlights the weaker QNMs of interest by removing stronger modes.  This approach has proven fruitful for theoretical waveforms obtained from numerical relativity \cite{Ma:2022wpv}. In this paper, a Bayesian framework allows us to apply these filters to real data, leading to more definite evidence for weaker QNMs. We will use GW150914 as an example for its application.



{\it The rational filter.---} In Ref.~\cite{Ma:2022wpv}, we showed that QNM(s) can be removed using the so-called ``rational filters''. Moreover, the presence of weaker modes in numerical-relativity ringdown waveforms, e.g., the second-order QNMs (also see \cite{Mitman:2022qdl,Cheung:2022rbm}) and retrograde modes, becomes more pronounced after removing the most dominant mode(s). Let us now clean a $\omega_{lmn}-$QNM from time-series data $d_t$ that contains both signal and noise.
We first transform $d_t$ to the frequency domain via fast Fourier transform (FFT), producing frequency-domain data $\tilde{d}_f$. Note that we typically take longer $d_t$ than the full IMR signal to avoid spectral leakage. We then apply a rational filter \cite{Ma_prd,Ma:2022wpv}
\begin{align}
    &\mathcal{F}_{lmn}(f;M_f,\chi_f)=\frac{2\pi f-\omega_{lmn}}{2\pi f-\omega_{lmn}^*}\frac{2\pi f+\omega_{lmn}^*}{2\pi f+\omega_{lmn}} ,\label{eq:rational_filter}
\end{align}
to $\tilde{d}_f$. Here the filter $\mathcal{F}_{lmn}$ implicitly depends on $M_f$ and $\chi_f$ --- the value of $\omega_{lmn}$ is fully determined by the two properties due to the no-hair theorem \cite{penrose2002golden,Chrusciel:2012jk,PhysRevLett.26.331,PhysRev.164.1776}. 
Finally, we transform the filtered frequency-domain data back to the time domain, given by
\begin{align}
    d^F_t=\int df \tilde{d}_f\mathcal{F}_{lmn}(f)e^{-i2\pi f t}.
\end{align}
Multiple QNMs can be cleaned simultaneously via a product filter:
\begin{align}
  \mathcal{F}_{\rm tot}=\prod_{lmn} \mathcal{F}_{lmn}. \label{eq:total_filter}
\end{align}
The filtered time-series data $d^F_t$ are still real-valued because each filter 
satisfies $\mathcal{F}_{lmn}(-f)=\left[\mathcal{F}_{lmn}(f)\right]^*$.

Since the filter $\mathcal{F}_{lmn}$ operates on the entire data segment, we list its impact on different parts of the signal and noise. For early inspiral, the filter introduces a trivial phase and negative time shift \cite{Ma:2022wpv}, which has no influence on the ringdown analysis. When using $\mathcal{F}_{lmn}$ to clean $\omega_{l m n}$, the amplitude of a different mode, $\omega_{l^\prime m^\prime n^\prime}$ $(\omega_{l^\prime m^\prime n^\prime}\neq \omega_{l m n})$, is reduced by a factor of $B_{lmn}^{l^\prime m^\prime n^\prime}$, given by \cite{Ma_prd}
\begin{align}
    B_{lmn}^{l^\prime m^\prime n^\prime} e^{i\varphi_{lmn}^{l^\prime m^\prime n^\prime}}\equiv\mathcal{F}_{lmn}(\omega_{l^\prime m^\prime n^\prime}). \label{eq:filter_on_other_qnm}
\end{align}
The symmetry of $\mathcal{F}_{lmn}$ yields an identity $B_{lmn}^{l^\prime m^\prime n^\prime}=B^{lmn}_{l^\prime m^\prime n^\prime}$.
The start time and oscillation of the $\omega_{l^\prime m^\prime n^\prime}$ mode remain unchanged. 
Finally, we emphasize that the filter has no impact on the statistical properties of the noise, including Gaussianity, stationarity, and the one-sided noise power spectral density (PSD), because $|\mathcal{F}_{lmn}(f)|=1$ for any real frequencies \cite{Ma_prd}.

{\it Integrating the filter into Bayesian inference.---} After cleaning enough QNMs with $\mathcal{F}_{\rm tot}$ in Eq.~\eqref{eq:total_filter}, we expect the ringdown portion of the filtered data $d^{F}_t$ to be consistent with pure noise. Thus we introduce a likelihood in time domain \cite{Ma_prd,Isi:2021iql}, 
\begin{align}
    \ln P\,(d_t|M_f,\chi_f,t_{0},\mathcal{F}_{\rm tot})=-\frac{1}{2}\sum_{i,j>I_0}d^{F}_iC_{ij}^{-1}d^F_j, \label{eq:log-p}
\end{align}
where $C_{ij}$ is the autocovariance function, $d^F_i\equiv d^{F}(t_i)$ denotes the samples of the filtered data after some truncation time $t_0$ $(t_i>t_0)$, and $I_0$ is the index associated with $t_0$. For any given $t_0$, the likelihood depends only on two parameters: $M_f$ and $\chi_f$, which are used to construct the filter. There is no additional dependence on the amplitudes and phases of QNMs --- the corresponding QNMs can be removed from the ringdown regime regardless of their amplitudes and phases. This is similar to the case where a constant can be removed by taking the derivative regardless of the constant value. 
In the Supplemental Material, we discuss the relation between our new approach and the usual full-ringdown (full-RD) MCMC analysis that does not apply the rational filter \cite{Isi:2021iql,Isi:2019aib}. Especially, we show that the rational filter eliminates the dependency on amplitudes and phases through a new maximum likelihood estimator, corresponding to an approximate marginalization over the mode amplitudes and phases.

Since Eq.~\eqref{eq:log-p} is merely a two-dimensional (2D) function, the direct evaluation of the likelihoods on a $M_f-\chi_f$ grid is efficient enough such that we do not need random sampling techniques, e.g., Markov chain Monte Carlo.
In addition, the likelihoods can be easily converted to the joint posteriors of $M_f$ and $\chi_f$ via
\begin{align}
    \ln P\,(M_f,\chi_f | d_t,t_{0},\mathcal{F}_{\rm tot}) &\propto \ln P\,(d_t|M_f,\chi_f,t_{0},\mathcal{F}_{\rm tot}) \notag \\
    &+ \ln \Pi\,(M_f,\chi_f), \label{eq:m_chi_posteriors}
\end{align}
where $\ln \Pi\,(M_f,\chi_f)$ is the prior.

\begin{figure}[htb]
    \includegraphics[width=\columnwidth,clip=true]{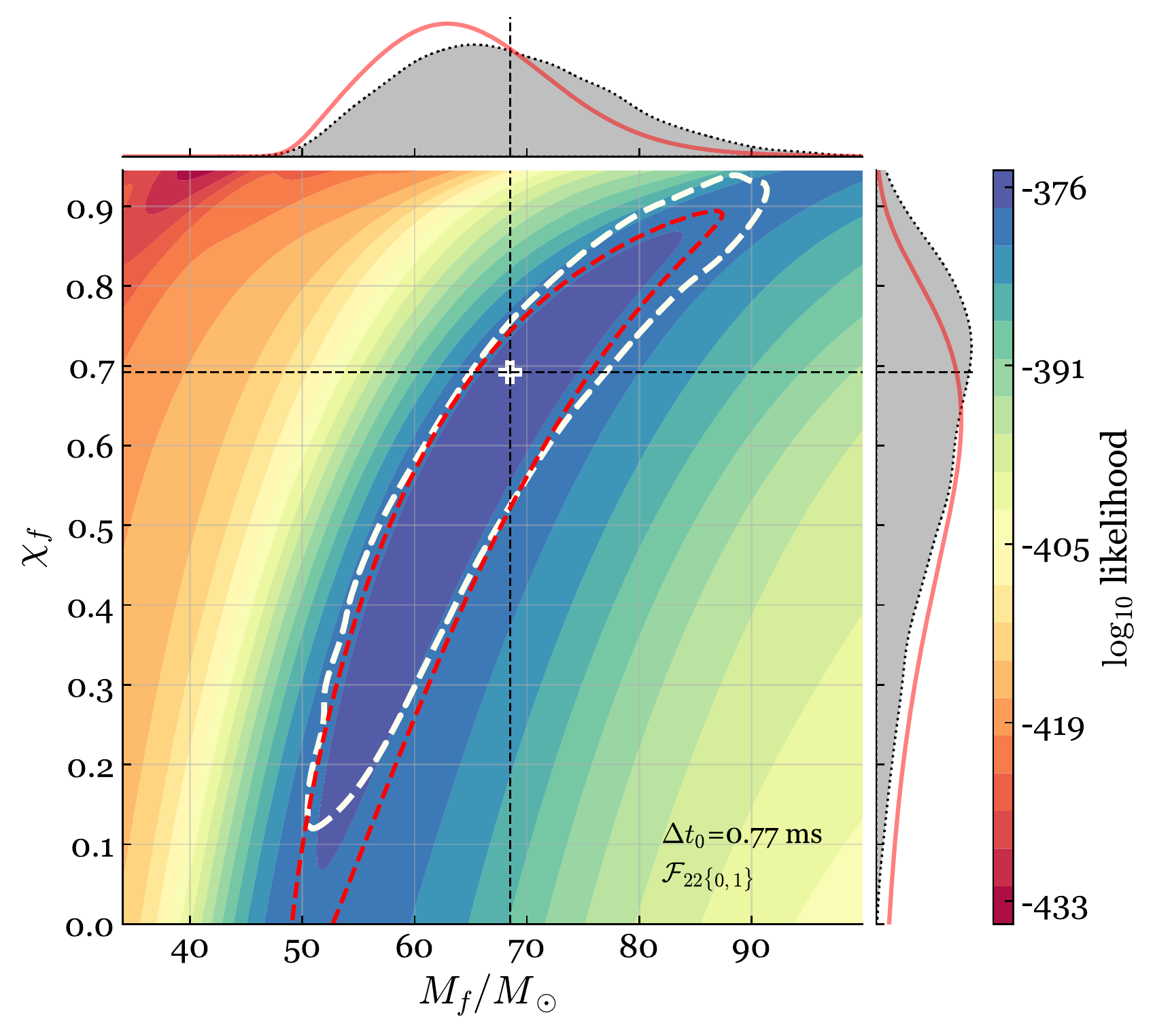}
  \caption{Joint posterior distribution of $M_f$ and $\chi_f$ for GW150914 ($\Delta t_0=0.77$ ms, $\mathcal{F}_{\rm tot}=\mathcal{F}_{221}\mathcal{F}_{220}$). We evaluate the posteriors on a 2D grid: $M_f\in\left[35M_\odot,100M_\odot\right]$ and $\chi_f\in[0,0.95]$, with a resolution of $\Delta M_f=0.1M_\odot$ and $\Delta \chi_f=5\times10^{-3}$. The red and white dashed contours indicate the 90\% credible region from the filtering and the full-RD MCMC analysis in \cite{Isi:2021iql,Isi:2019aib}, respectively.
  The plus sign marks the full IMR estimates $(M_f^{\rm IMR}=68.5M_\odot,\chi^{\rm IMR}_f=0.69)$. The 1D marginal distributions obtained from filtering (red) and the conventional full-RD MCMC analysis (gray) are shown in side panels.}
 \label{fig:GW150914_m_chi_novel_posteriors}
\end{figure}

We demonstrate the analysis using the LIGO Hanford and Livingston data of GW150914 \cite{Abbott:2016blz,LIGOScientific:2018mvr,GW_open_science_center}. We set the inferred GPS time at geocenter when the signal strain reaches the peak, $t_{\rm peak} = 1126259462.4083$ \cite{Isi:2019aib}, and parameterize the analysis starting time via \mbox{$\Delta t_0=t_0-t_{\rm peak}$}. We use the $\texttt{PYTHON}$ package $\texttt{ringdown}$ \cite{Isi:2021iql,ringdown_isi} to condition the data: (a) remove contents below 20 Hz via a high-pass filter and (b) downsample the data to 2048 Hz. We use the Welch method \cite{1161901} to estimate the PSDs of the two detectors with a 32-s segment of the conditioned data. The PSDs are then converted to the autocovariance function $C_{ij}$. We align the signals at the two detectors using the sky position of this event \cite{Isi:2019aib}: right ascension $\alpha=1.95$ rad and declination $\delta=-1.27$ rad. The information about the polarization and inclination angles is not needed. We fix the width of the ringdown analysis window to 0.2 s.

The joint posteriors of $M_f$ and $\chi_f$ evaluated with Eq.~\eqref{eq:m_chi_posteriors} are shown in Fig.~\ref{fig:GW150914_m_chi_novel_posteriors}. We set $\Delta t_0=0.77\,{\rm ms}$ and choose flat priors on a 2D grid. Here $\Delta t_0=0.77\,{\rm ms}$ is chosen such that the time is late enough for the signal to be consistent with a pure ringdown (Fig.~14 of \cite{Ma_prd}) and early enough for the first overtone to show its significance (Fig.~8 of \cite{Ma_prd}). We clean both the fundamental mode and the first overtone using the product filter $\mathcal{F}_{\rm tot}=\mathcal{F}_{221}\mathcal{F}_{220}$. With the posteriors computed over the whole parameter space, we calculate the 90\% credible region (red contour) by integrating the posteriors. The 1D distributions of $M_f$ and $\chi_f$ [by marginalizing Eq.~\eqref{eq:m_chi_posteriors}] are plotted as red curves in the side panels. For comparison, we repeat the usual full-RD MCMC analysis in Ref.~\cite{Isi:2021iql} using the $\texttt{ringdown}$ package \cite{ringdown_isi}. The white contour shows the 90\% credible region of the joint posteriors from the full-RD analysis. The joint distribution is also projected to the subspace of $M_f$ and $\chi_f$ (gray shades). The white plus sign marks the remnant mass and spin inferred from the full IMR analysis $(M_f^{\rm IMR}=68.5M_\odot,\chi^{\rm IMR}_f=0.69)$ \cite{Isi:2019aib}. Note that the mass is measured in the detector frame; the corresponding source-frame mass is $\sim62M_\odot$ \cite{LIGOScientific:2016aoc}. Studies at various $\Delta t_0$ times are discussed in our companion paper \cite{Ma_prd}. All the results are consistent with the conventional full-RD MCMC analysis \cite{Isi:2021iql}.


\begin{figure}[htb]
        \includegraphics[width=\columnwidth,clip=true]{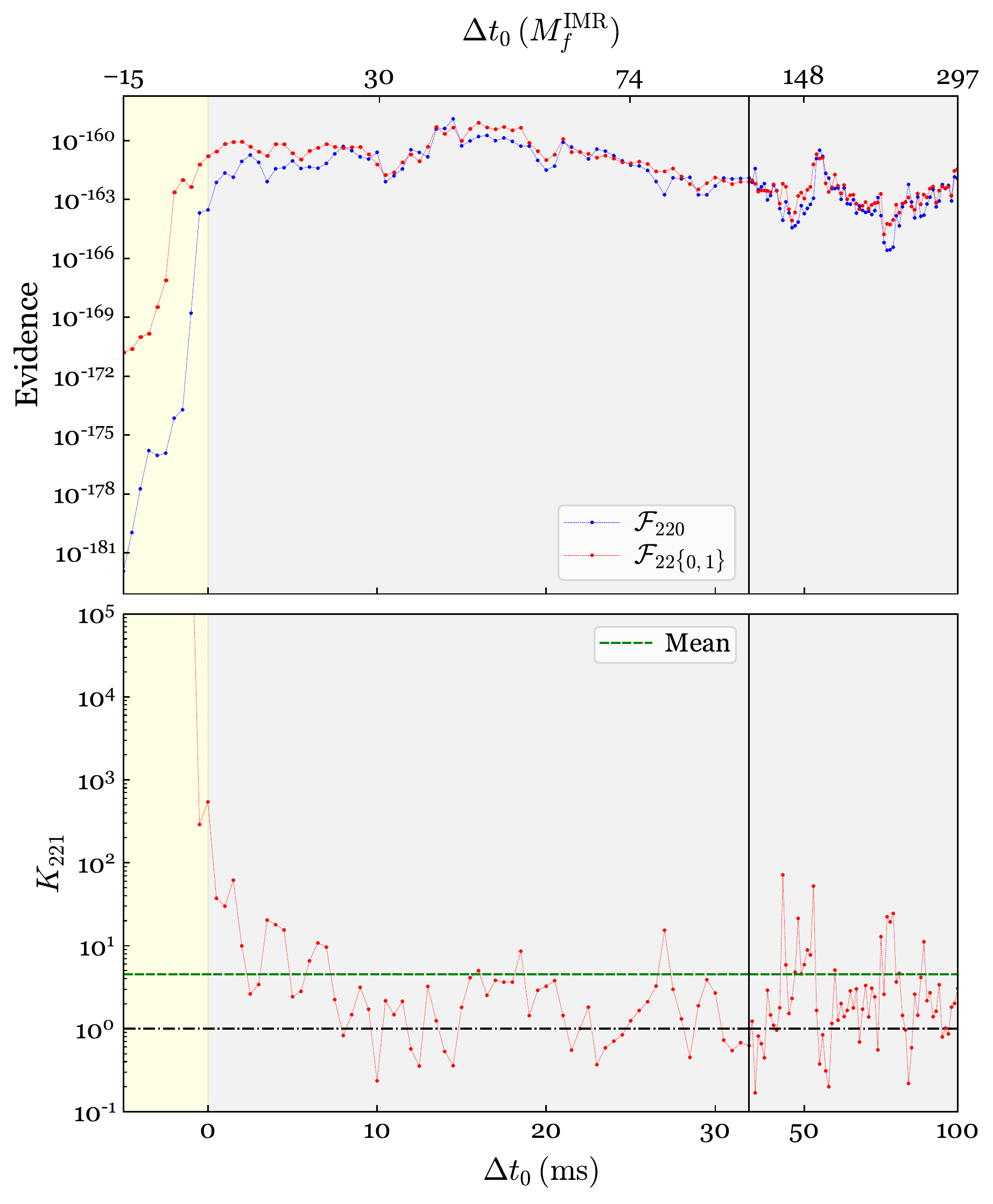}
  \caption{Model evidence (top) and Bayes factor (bottom) as a function of $\Delta t_0$. 
  Top: the red (blue) curve indicates the results with (without) the first overtone removed.
  In the bottom panel, the ratio between the two is shown as the Bayes factor [Eq.~\eqref{eq:bayes_overtone}]. The green dashed line indicates mean $K_{221}$ over $\Delta t_0\in [15,100]\, {\rm ms}$. The black dash-dotted line marks unity.}
 \label{fig:noval_likelihood_time}
\end{figure}

\begin{figure}[htb]
        \subfloat[Fundamental Mode\label{fig:mode_amplitudes_fun}]{\includegraphics[width=\columnwidth,clip=true]{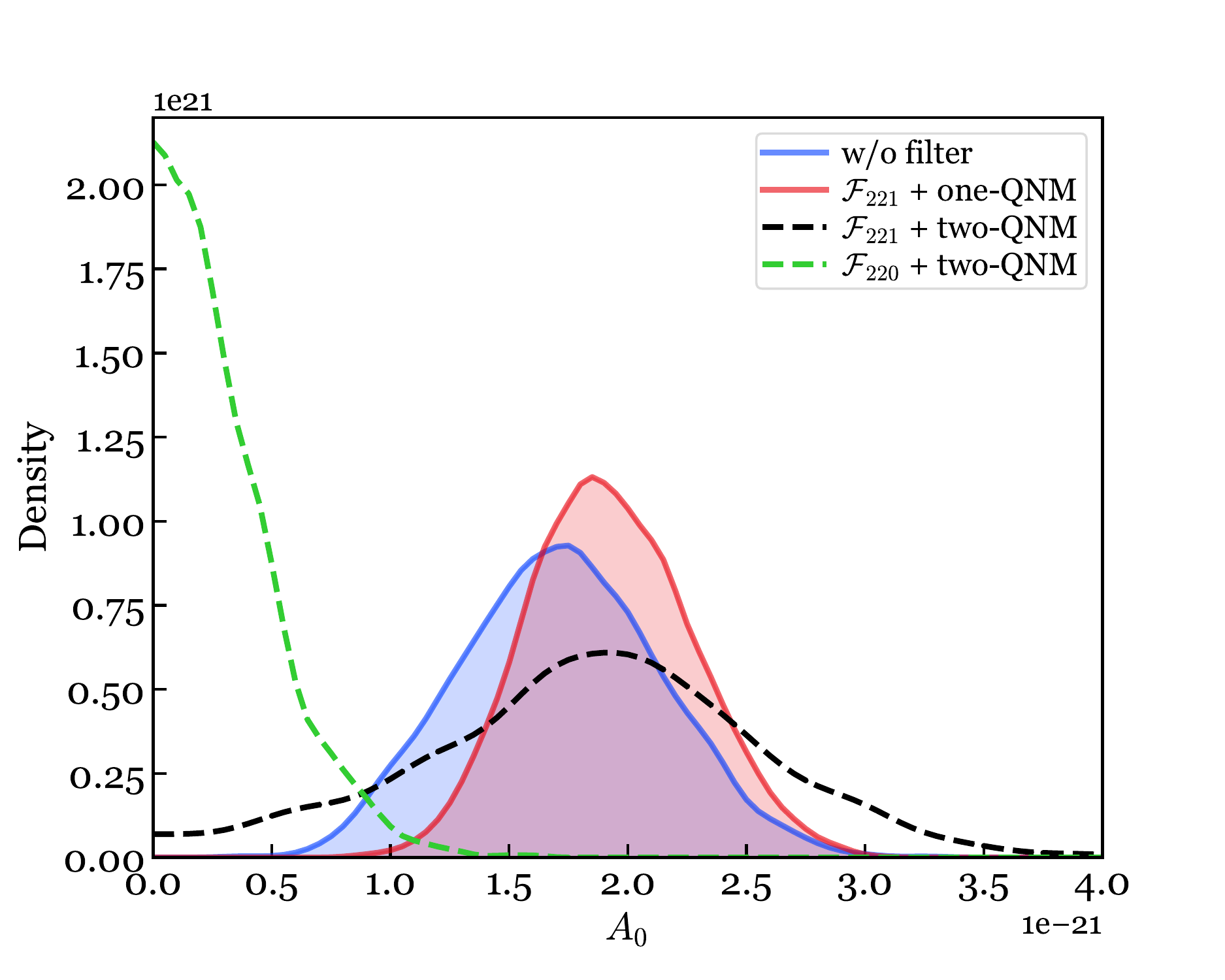}}\\
        \subfloat[First Overtone\label{fig:mode_amplitudes_overtone}]{\includegraphics[width=\columnwidth,clip=true]{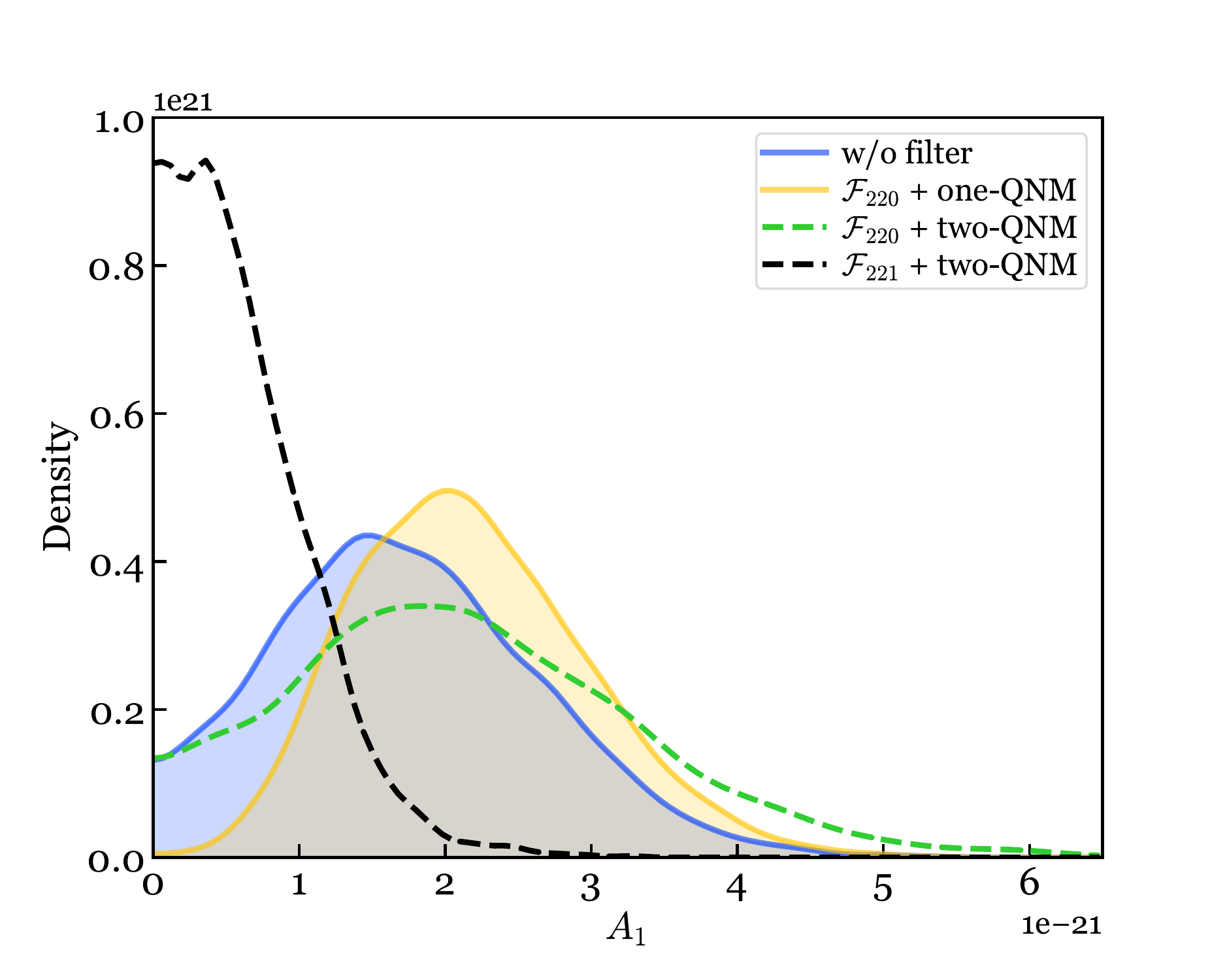}} 
  \caption{Posteriors of the fundamental (top) and first overtone (bottom) mode amplitudes in GW150914 ($\Delta t_0=0.77$ ms). The blue regions are obtained from the full-RD MCMC analysis (without filters). After cleaning the fundamental mode, the filtered data are fitted with a two-QNM template (green) and an overtone-only template (yellow). After removing the first overtone, the data are fitted with a two-QNM template (black) and a fundamental-mode-only template (red). Throughout these analyses we always place a uniform prior on $A_{0,1}$.}
 \label{fig:mode_amplitudes}
\end{figure}

{\it Model evidence and Bayes factor.---} We compute the model evidence by integrating the likelihoods in Eq.~\eqref{eq:log-p}
\begin{align}
P(d_t|\Delta t_0,\mathcal{F}_{\rm tot})=&\iint P\,(d_t|M_f,\chi_f,\Delta t_{0},\mathcal{F}_{\rm tot})\notag \\
&\times\Pi(M_f,\chi_f)dM_fd\chi_f. \label{eq:filter_prob}
\end{align}
Assuming a flat prior, the top panel of Fig.~\ref{fig:noval_likelihood_time} shows the evidence as a function of $\Delta t_0$, using $\mathcal{F}_{\rm tot}=\mathcal{F}_{220}$ (blue) and $\mathcal{F}_{\rm tot}=\mathcal{F}_{221}\mathcal{F}_{220}$ (red). Both curves surge quickly around $\Delta t_0\sim0$, implying the onset of the ringdown regime, and then level off, followed by oscillations around a plateau. Both curves start to rise before $\Delta t_0=0$ because the width of the analysis window $0.2$ s is much wider than the ringdown signal; the full ringdown is already in the window even when $\Delta t_0$ is still slightly negative. The evidence increases sharply as the inspiral signal slides off the ringdown window. Such a generic trend of the evidence around a ringdown signal naturally offers us an agnostic estimate of the start time of a QNM.

In addition, we calculate the Bayes factor by taking the ratio of the two evidence curves: 
\begin{align}
    K_{221}(\Delta t_0)=\frac{P(d_t|\mathcal{F}_{22\{0,1\}},\Delta t_0)}{P(d_t|\mathcal{F}_{220},\Delta t_0)}. \label{eq:bayes_overtone}
\end{align}
The results are plotted in the bottom panel of Fig.~\ref{fig:noval_likelihood_time}. At $\Delta t_0\lesssim 2\,{\rm ms}$, $K_{221}$ decreases sharply. Later, it fluctuates around a mean value of 4.5 for $\Delta t_0\in [15,100]\, {\rm ms}$, when we expect no ringdown signal remains. We find a Bayes factor as high as 600 at the peak time $\Delta t_0=0$, demonstrating a strong preference for the existence of the first overtone.

{\it MCMC analysis after mode cleaning.---} Detecting a subdominant QNM with the conventional MCMC approach (without mode cleaning) is complicated by the presence of the dominant mode(s), especially at a low-SNR regime. 
We can take advantage of mode cleaning to surmount such challenges.  
We first use the rational filter to remove the dominant mode(s) and then use the established criteria (in terms of the standard MCMC) to look for the evidence of the subdominant QNM(s) in the filtered data. This forms a hybrid approach for BH spectroscopy.

We now demonstrate the hybrid approach using GW150914 ($\Delta t_0=0.77\,{\rm ms}$), where we always place uniform priors on the mode amplitudes. In Fig.~\ref{fig:mode_amplitudes}, the posteriors of the mode amplitudes from the unfiltered data with the full-RD MCMC method \cite{Isi:2021iql,ringdown_isi} are shown in blue. After cleaning the fundamental mode, with $M_f$ and $\chi_f$ fixed to the IMR-estimated values, we fit the filtered data using MCMC with a two-QNM ringdown template including both the fundamental mode and the first overtone, i.e., assuming we are agnostic of mode cleaning. The results are shown as green dashed curves in Fig.~\ref{fig:mode_amplitudes}. The posterior of the fundamental mode amplitude is $\sim 0$. The distribution of the first overtone amplitude remains unchanged (consistent with the blue distribution). Note that since the filter reduces the amplitude of the first overtone by $B_{220}^{221}=0.487$ [Eq.~\eqref{eq:filter_on_other_qnm}], we compensate for this factor in the figure for a fair comparison. The results demonstrate the successful removal of the fundamental mode. We also fit the filtered data with a one-QNM, overtone-only template, leading to the yellow distribution in Fig.~\ref{fig:mode_amplitudes_overtone}, consistent with both the two-QNM fitting results (green) and the unfiltered data fitting results (blue).

\begin{figure}[htb]
        \includegraphics[width=\columnwidth,clip=true]{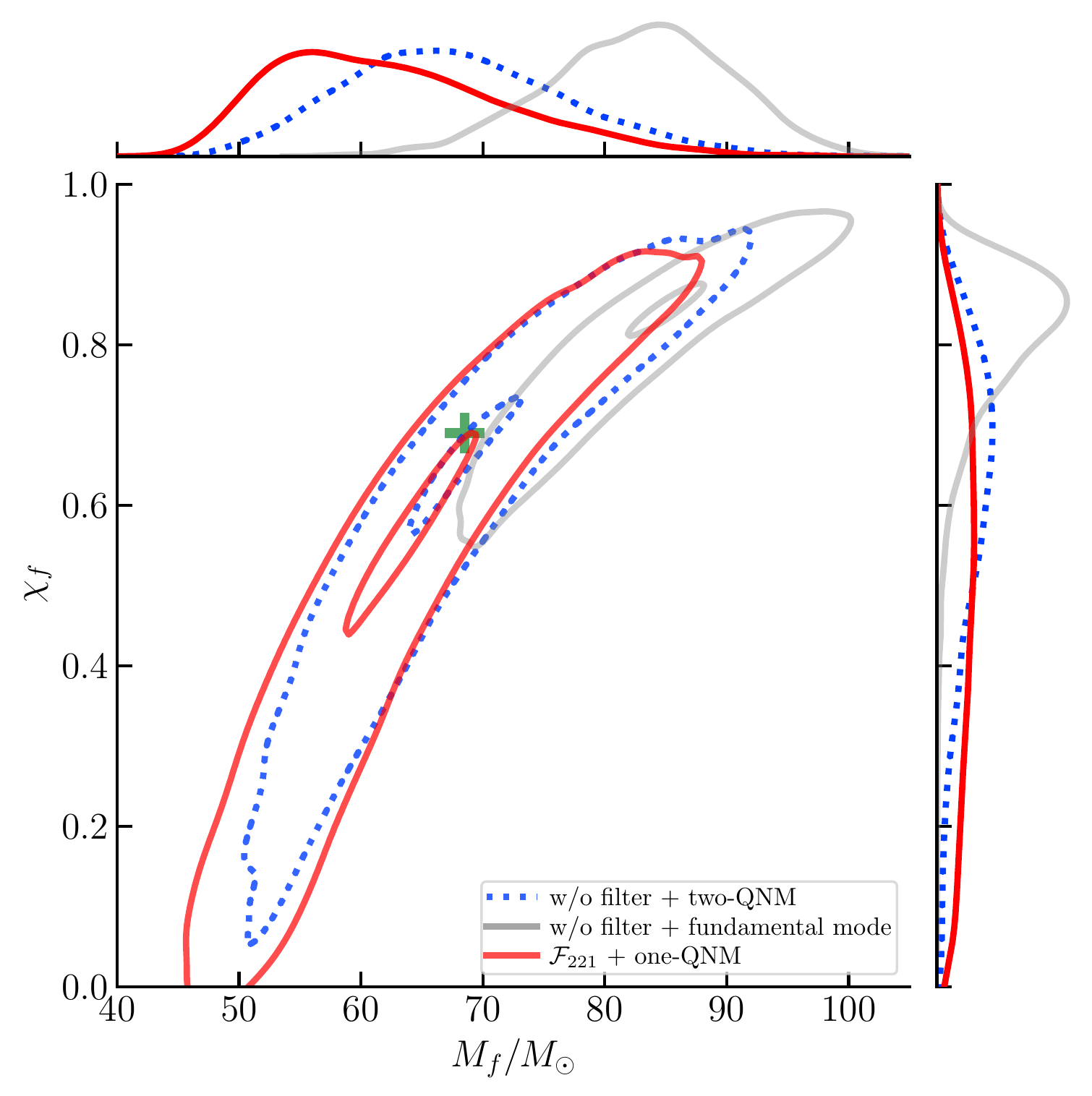}
  \caption{Posteriors of $M_f$ and $\chi_f$ inferred from the ringdown of GW150914 ($\Delta t_0=0.77\,{\rm ms}$). 
  The full-RD MCMC approach (without filters) yields the blue dashed contours using a two-QNM model ($\omega_{220}$ and $\omega_{221}$) and the gray contours using a fundamental-mode-only template.
  After cleaning the first overtone, fitting the filtered data using a fundamental-mode-only template leads to the red contours. The green plus sign marks the IMR-estimated values.}
 \label{fig:mchi_ligo_prl}
\end{figure}

Similarly, we repeat the procedure to remove the first overtone. The black dashed curves and the red region correspond to fitting the filtered data with a two-QNM and a fundamental-mode-only template, respectively. This time, the black curve indicates the removal of the first overtone, whereas the fundamental mode is generally not impacted.

Next, we show the estimates of $M_f$ and $\chi_f$ for GW150914 at $\Delta t_0=0.77$ ms using the hybrid approach (Fig.~\ref{fig:mchi_ligo_prl}). Here we clean the first overtone and then fit the filtered data with the fundamental-mode-only template. The resulting constraints are shown in red, consistent with the IMR-estimated values (green plus sign) and the results obtained from the full-RD MCMC analysis with a two-QNM template (blue). On the contrary, the results are biased when we fit the unfiltered data with the fundamental mode alone (gray), i.e., with solely the fundamental mode, the model is not good enough to describe the ringdown at $\Delta t_0=0.77$ ms.

\begin{figure}[htb]
        \includegraphics[width=\columnwidth,clip=true]{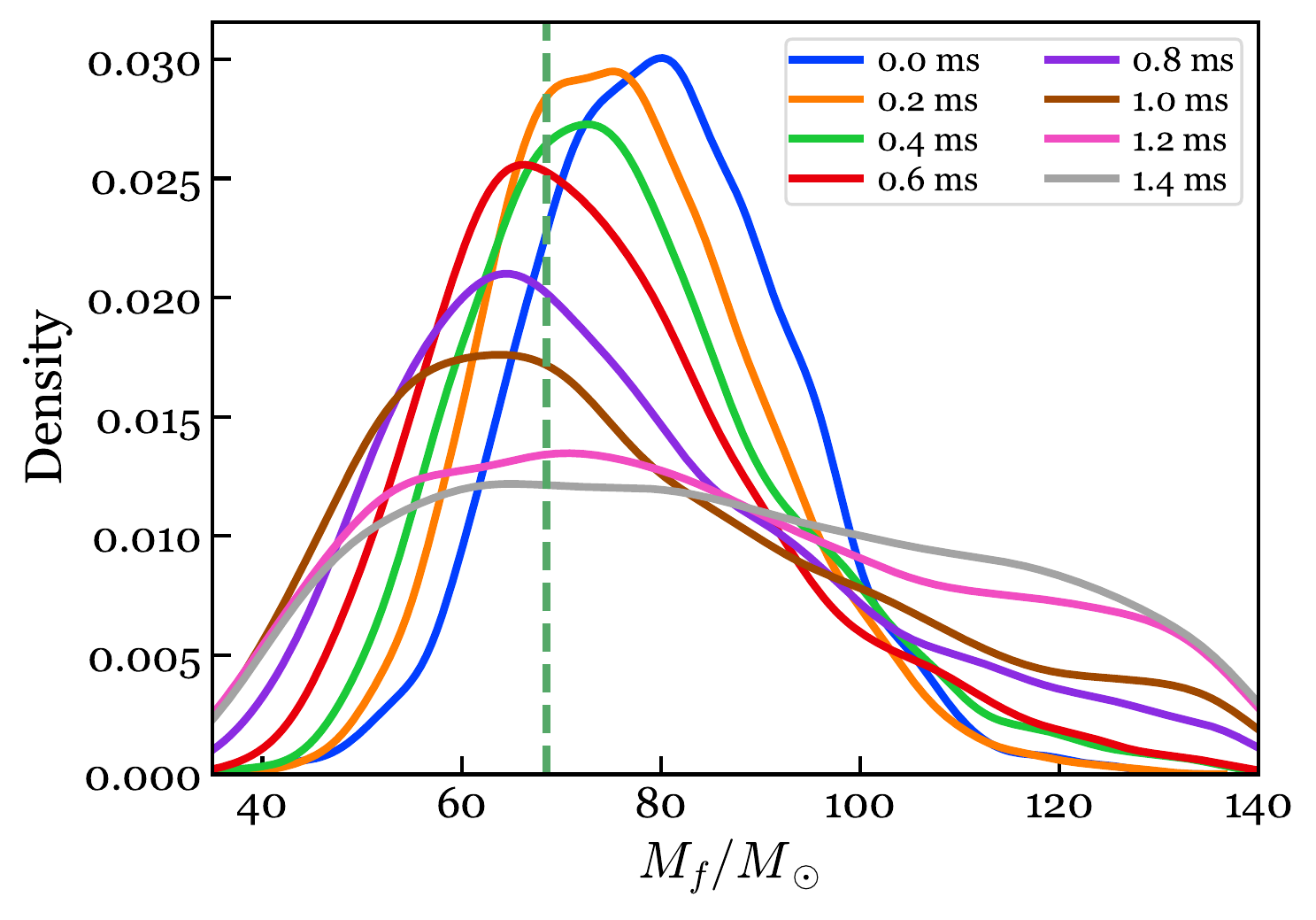}
  \caption{Posteriors of $M_f$ inferred from the first overtone alone at different starting times. We clean the fundamental mode using $\mathcal{F}_{220}$ and fit the filtered data with the first-overtone-only template. We place a uniform prior in the range shown on the $x$-axis: $M_f \in [35M_\odot, 140M_\odot]$. The vertical line indicates the IMR-estimated mass. The colors of the curves correspond to different $\Delta t_0$ times (see legend). See Fig.~16 of \cite{Ma_prd} for the equivalent plot for $\chi_f$; the joint posteriors of $M_f$ and $\chi_f$ are in Fig.~14 (c) and (d) of \cite{Ma_prd}.}
 \label{fig:no_hair}
\end{figure}

Finally, we present the posteriors of $M_f$ inferred from the first overtone alone at different $\Delta t_0$ times (Fig.~\ref{fig:no_hair}). We set uniform prior in the range of $35M_\odot\leq M\leq 140M_\odot$. At the signal peak ($\Delta t_0=0$), the maximum a posteriori (MAP) value of $M_f$ is higher than the IMR-estimated value (vertical line), which may indicate the existence of residual signals in addition to the two QNMs ($\omega_{22,0\&1}$); see more details in \cite{Ma_prd}. At later times, the MAP value shifts toward smaller $M_f$ and becomes more consistent with the IMR value. In the end, the distribution flattens and becomes less informative at $\Delta t_0\gtrsim1\,{\rm ms}$ when most of the overtone signal disappears.

{\it Discussions.---} In this paper, leveraging the rational filters~\cite{Ma:2022wpv}, we  
formulate a mode-cleaning-based Bayesian framework for BH spectroscopy.
More details are provided in \cite{Ma_prd}. Some of our calculations are implemented in a public $\texttt{PYTHON}$ package, {\texttt{qnm\_filter}} \cite{QNMfilter}. Complementary to the existing time-domain \cite{Carullo:2019flw,Isi:2021iql} and frequency-domain studies \cite{Finch:2021qph}, our new framework has a few advantages. 

(a) The rational filters offer a simple way to remove particular QNM(s) from the ringdown data. After cleaning the most dominant mode(s), the inferences for a subdominant mode become more definitive, especially when its SNR is low.

(b) The filter adopts a new maximum likelihood estimator to eliminate the dependency on mode amplitudes and phases (an approximate marginalization). The resulting likelihoods depend only on the remnant BH's mass and spin. This feature allows a fully parallelizable pipeline to efficiently evaluate the likelihoods for a wide range of starting times without MCMC, free of technical issues like convergence.

(c) 
Model evidence is presented as a function of analysis start time.
Although the signal peak and the inferred start of ringdown can be obtained from the full IMR analysis, the estimate might degrade when the system configurations and parameters go beyond the available regime of the corresponding IMR waveform model, or when the inspiral signal is too short. The information obtained from the evidence rising time could contribute to the prior distribution of the start time in other frameworks, e.g., Refs.~\cite{Carullo:2019flw,Finch:2021qph}. On the other hand, we adopt the fixed sky position throughout this analysis to align the data from two LIGO detectors. The analysis may be biased if the estimated sky position is inaccurate. Given that the rising time of the evidence curve provides an independent estimate of the ringdown start time, an alternative is to apply a ringdown-driven time lag between data from two detectors based on the alignment of the evidence curves. We leave relevant studies for future work.

This work is the first demonstration of an overtone mode in GW150914 after cleaning the fundamental mode. Future application of our framework to other events, e.g., GW190521 \cite{Capano:2021etf,Capano:2022zqm}, and new events in the forthcoming fourth observing run, may lead to discovery and evidence of other modes in BH ringdown. 
As demonstrated with numerical-relativity waveforms \cite{Ma:2022wpv}, the rational filters provide a powerful tool to reveal subdominant effects, e.g., second-order nonlinearity and retrograde modes. Future detectors will enable us to do BH spectroscopy in a high-SNR regime. 
Our framework will allow us to test general relativity and explore the nature of gravity in full detail using real events.

\begin{acknowledgments}
{\it Acknowledgments.---}
We thank Maximiliano Isi and Will M. Farr for their helpful suggestions and comments. We also thank Eric Thrane, Paul Lasky, Emanuele Berti, Mark Ho-Yeuk Cheung, Roberto Cotesta, and all the attendees of the ringdown workshop held at CCA, Flatiron Institute for useful discussions. Computations for this work were performed with the Wheeler cluster at Caltech. YC and SM acknowledge support from the Brinson Foundation, the  Simons Foundation (Award Number 568762), and by NSF Grants PHY-2011961, PHY-2011968, PHY--1836809.
LS acknowledges the support of the Australian Research Council Centre of Excellence for Gravitational Wave Discovery (OzGrav), Project No. CE170100004. This material is based upon work supported by NSF's LIGO Laboratory which is a major facility fully funded by the National Science Foundation.
\end{acknowledgments}


\def\bibsection{\section*{References}}
\bibliography{References}
\end{document}